\documentclass[letterpaper,10pt]{article}
\usepackage{osameet2}
\usepackage{graphicx}

\usepackage{amsmath,amssymb}
\usepackage{soul}

\usepackage[colorlinks=true,bookmarks=false,citecolor=blue,urlcolor=blue]{hyperref} %latex w/dvips

\newcommand{\nft}{\textnormal{NFT}}
\newcommand{\inft}{\textnormal{INFT}}
\renewcommand\refname{5.~~~ References}

\begin{document}

\title{Nonlinear Frequency-Division Multiplexing in the Focusing Regime}

\author{Xianhe Yangzhang\textsuperscript{(1)}, Mansoor I. Yousefi\textsuperscript{(2)}, 
Alex Alvarado\textsuperscript{(3)}, Domanic Lavery\textsuperscript{(1)}, Polina Bayvel\textsuperscript{(1)}}

\address{\textsuperscript{(1)} Department of Electronic and Electrical Engineering, University College London, UK\\
	\textsuperscript{(2)} Communications and Electronics Department, T\'el\'ecom ParisTech, Paris, France\\
	\textsuperscript{(3)} Department of Electrical Engineering, Eindhoven University of Technology, The Netherlands
	}

\begin{abstract}
Achievable rates of the nonlinear frequency-division multiplexing (NFDM) and wavelength-division multiplexing (WDM) subject to the same
power and bandwidth constraints are 
computed as a function of transmit power in the standard single-mode fiber. NFDM achieves higher rates than WDM.
\end{abstract}

\ocis{060.2330, 060.4370}

\section{Introduction}

The achievable rates of the linear and nonlinear frequency-division
multiplexing (NFDM) were compared in fibers 
with normal dispersion in \cite{yousefi2016nfdm}. In this case, the discrete eigenvalues are naturally absent in the
nonlinear Fourier transform (NFT). It was shown that the NFDM rate
increases monotonically with 
transmit power in simulations, while the rates of the wavelength-division multiplexing (WDM) characteristically 
vanish at an optimal power.

In this paper, we generalize the results of \cite{yousefi2016nfdm} from normal dispersion fiber (defocusing regime,
with negative dispersion $D$) to the anomalous dispersion fiber (focusing regime, with positive dispersion $D$).   
In the focusing regime, solitons can be present. However, we set the discrete spectrum to zero and modulate merely the
continuous spectrum in the NFT. We
assume a network scenario, defined formally in \cite[Section II. A]{yousefi2016nfdm}, and compare the achievable rates of the
NFDM and WDM subject to the same power and bandwidth constraints. 

The continuous spectrum modulation is studied in \cite[Part
III, Section V. D]{part123}, \cite{prilepsky2013,le2014nis,wahls2015digital,aref2016ecoc,tavakkolnia2015sig}. However, nonlinear signal 
multiplexing, the main ingredient in NFDM responsible for high data rates, was not realized until 
recently in the defocusing regime, where the NFT simplifies
\cite{yousefi2016nfdm}. Following 
\cite{yousefi2016nfdm}, the authors of \cite{turitsyn2016nature}
considered nonlinear multiplexing in multi-user systems in the
focusing regime. However, computation of the spectral efficiency (SE), taking into account time duration and bandwidth, 
and its comparison with the SE of the WDM under the same constraints is lacking. In this paper, we undertake 
such study.

In Section~\ref{sec:nfdm}, we recall a few equations from \cite{yousefi2016nfdm} that are used in this
paper. The achievable rates of NFDM and WDM are computed in Section~\ref{sec:nfdm}. In Section~\ref{sec:comparison}, 
we comment on  nonlinear multiplexing in the focusing and de-focusing regimes.

\section{Nonlinear Frequency-Division Multiplexing}
\label{sec:nfdm}

Signal propagation in single-mode optical fiber with distributed amplification is modeled by the 
stochastic nonlinear Schr\"odinger (NLS) equation. The equation can be normalized as \cite[Eq. 3]{part123}
\begin{equation*}
	j\frac{\partial q}{\partial z}= \frac{\partial ^2 q}{\partial t^2}+2\vert q\vert^2q+n(t,z),\ \ 0\leq z\leq\mathcal{L},
\end{equation*}
where $q(t,z)$ is the complex envelope of the signal as a function of time $t$ and distance $z$ and
$n(t,z)$ is zero-mean white circularly-symmetric Gaussian noise.

Consider a multi-user system with $N_u$ users, each sending $N_s$ symbols, with total (linear) bandwidth
$B$ Hz and (total) average power $\mathcal P$.   
Let $\lambda\in\mathbb R$ be the generalized frequency, in Fourier transform relation with the 
generalized time $\tau$. The NFDM transmitter (TX) consists of three
steps. In the first step, the following function is computed 
\begin{equation}
	u(\tau,0)=\sqrt{2}\sum_{k=-\frac{N_u}{2}}^{\frac{N_u}{2}-1}\sum_{\ell=-\frac{N_s}{2}}^{\frac{N_s}{2}-1}
s^k_{\ell}\phi(\tau-\ell T_0)e^{j2\pi kW_0\tau},
%=\sqrt{2}\mathcal{F}^{-1}(U(\lambda,0)),
\label{eq:modulation}
\end{equation}
where $\phi(\tau)$ is a root-raised-cosine function with the bandwidth
$W_0$ Hz and the roll-off factor $r$, $T_0=1/W_0$, and $\{s_l^k\}_{l}$ are symbols of user $k$ chosen from a constellation $\Xi$.

In the second step, the transmit signal in the nonlinear Fourier domain is calculated
 \begin{equation*}
	\hat{q}(\lambda,0)=\Bigl(e^{\vert U(\lambda,0)\vert^2}-1\Bigr)^{\frac{1}{2}}e^{j\angle U(\lambda,0)},
\end{equation*}
in which $U(\lambda, 0)=\mathcal F(u(\tau,0))/\sqrt 2$, where $\mathcal
F$ denotes Fourier transform. Finally, $q(t,0)=\inft(\hat q(\lambda,0))$.

The NFDM receiver (RX) similarly consists of three steps. In the first step, the forward NFT is applied to obtain 
$\hat q(\lambda, \mathcal L)=\nft(q(t,\mathcal L))$. In the second
step, first channel equalization is performed 
\begin{equation*}
	\hat{q}(\lambda,0)=\hat q(\lambda,\mathcal{L}) e^{4j\lambda^2\mathcal{L}};
\end{equation*}
then the $U$ function is computed  
\begin{equation*}
U(\lambda,\mathcal L)=\Bigl(
\log\bigl(1+\vert\hat{q}(\lambda,0)\vert^2\bigr)
\Bigr)^{\frac{1}{2}}e^{j\angle \hat q(\lambda,0)}.
\end{equation*}
In the third step, received symbols are obtained by match filtering
 \begin{equation*}
	s^k_{\ell}=\int_{-\infty}^{\infty}u(\tau,\mathcal{L})\phi^*(\tau-\ell T_0)e^{-2\pi jkW_0\tau} d\tau,
\end{equation*}
where $u(\tau,\mathcal L)=\sqrt 2\mathcal F^{-1}(U(\lambda,\mathcal L))$.

Constellation $\Xi$ and the nonlinear bandwidth $W_0$ are chosen so that the linear bandwidth of $q(t,z)$ is $B$ Hz.

\section{Achievable Rates of the NFDM and WDM}
\label{sec:rates}

We consider $N_u=5$, $N_s=1$, $B=100$ GHz, $\mathcal L=2000$ km, $r=25\%$ and fiber 
parameters in \cite[Part III, Table~I]{part123}. One nonlinear multiplexer at the input, 
and one nonlinear de-multiplexer at the output,  are implemented  in the NFDM. The same setup is considered
for WDM. The equalization in WDM is applying back-propagation to the central user.  
In this paper, bandwidth and time duration of the signals are defined as the 
intervals that contain 99\% of the signal energy. 

\begin{figure}[!htb]
\centerline{
\begin{tabular}{ccc}
\includegraphics[width=0.33\textwidth]{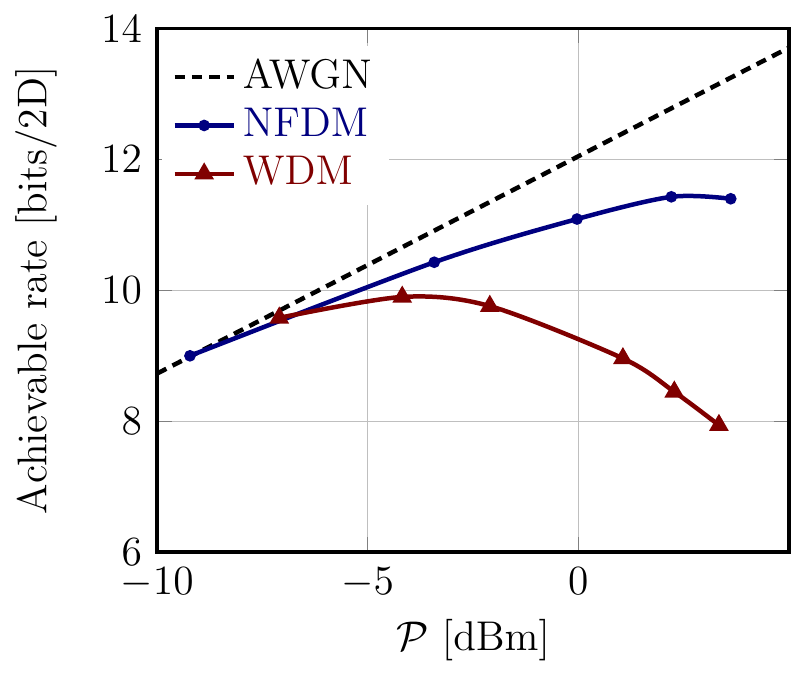}
&
\includegraphics[width=0.3\textwidth]{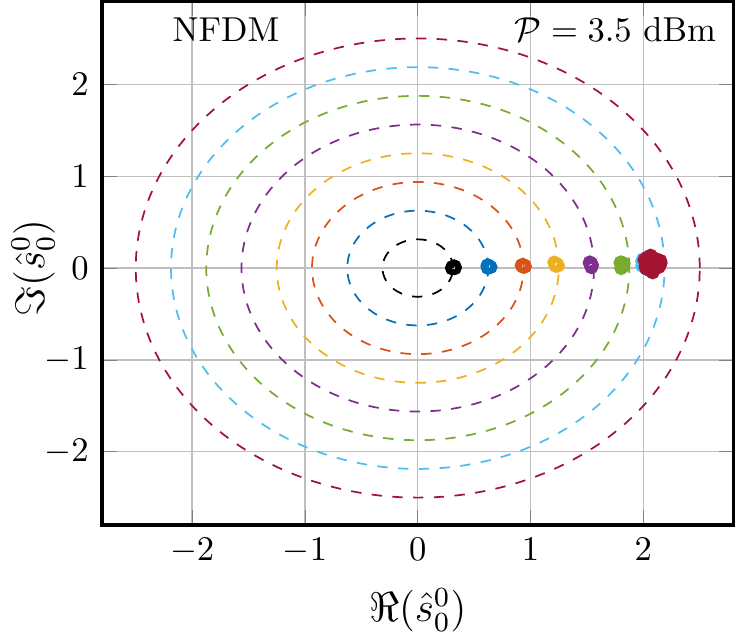}
&
\includegraphics[width=0.3\textwidth]{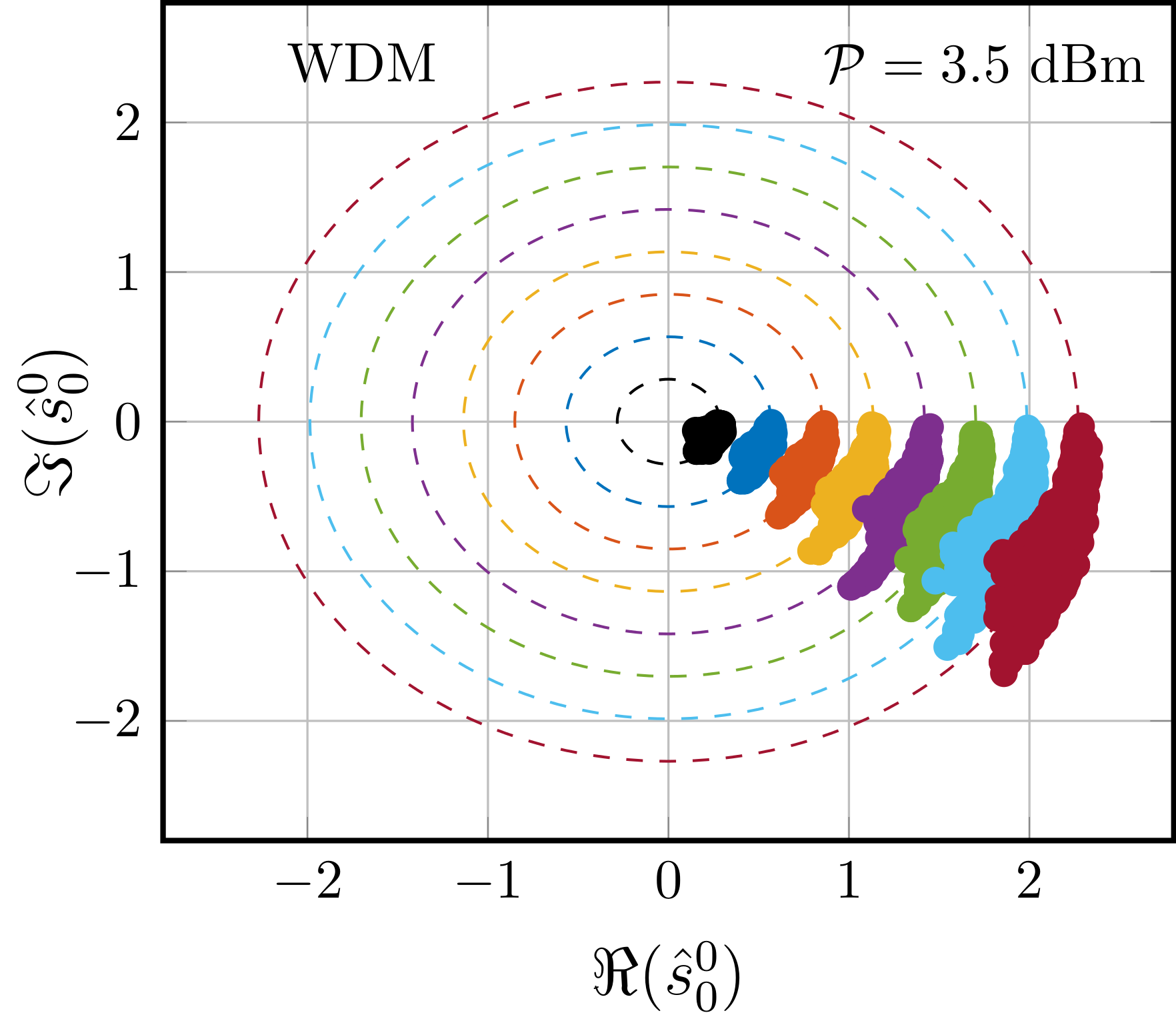}
\\
~~~~(a) & ~~~~~~~(b) &~~~~~~~(c)
\end{tabular}
}
\caption{ (a) Achievable rates of the NFDM and WDM. Transmit and
  received symbols in (b) NFDM and (c) WDM, at the same power $\mathcal P=3.5$ dBm.}
\label{fig:rates}
\end{figure}

The constellation $\Xi$ consists of $32 < N_r <64$ rings with $N_{\phi}=128$ phase points in each ring. The number of samples in time, 
linear and nonlinear frequency is $N=2^{14}$. We compute the transition probabilities for the central 
symbol in the central user $s^0_0 \rightarrow \hat{s}^0_0$ based on 4500 noise realizations. The mutual information is maximized using 
the Arimoto-Blahut algorithm. The resulting rates as a function of the launch power are plotted in 
Fig.~\ref{fig:rates} (a). It can be observed that, while the WDM rates rolls off near
$\mathcal P=-4.5$ dBm, the NFDM increases for the range of power shown
in Fig.~\ref{fig:rates} (a). 

The data rates shown in  Fig.~\ref{fig:rates} (a) are measured in bits
per two real dimensions (bits/2D). They can be converted to the
spectral efficiency (SE) by dividing bits/2D by
$\alpha= B_{\max}T_a/(N_uN_s)$, where $T_{a}$ is the average time duration of the signal set at the 
input and $B_{\max}$ is the maximum input output  bandwidth. Asymptotically as the transmission 
time tends to infinity, $\alpha\rightarrow 1$ so that the rates shown in Fig.~\ref{fig:rates} (a) 
represent SE.  In the finite-blocklength regime with $N_s=1$, to give
an example, the SE of the NFDM and WDM at $\mathcal P=3.5$ dBm are,
respectively, 1.81 and 1.31 bits/s/Hz

Note that the signal power defined based on the $99\%$ time duration
may not be
a good definition for capacity. The rates shown in Fig.~\ref{fig:rates} (a) could change using
different signals with the same power. Nevertheless,
$\max R_{\textnormal{WDM}}(\mathcal P)<\max R_{\textnormal{NFDM}}(\mathcal P)$. More work is needed for a
comprehensive comparison.

\section{Stochastic Memory}
\label{sec:comparison}

Consider the mutual information for the central user $s_0^0\mapsto \{\hat s_l^0\}_l$. Using the chain 
rule for the mutual information   
\begin{equation}
I\bigl(s_0^0; \{\hat s_l^0\}_l\bigr)=I\bigl(s_0^0; \hat
s_0^0\bigr)+\underbrace{I\bigl(s_0^0; \{\hat s_l^0\}_{l\neq 0}\vert
  \hat s_0^0\bigr)}_
{\textnormal{stochastic memory}}.
\label{eq:chain-rule}
\end{equation}
In the absence of noise there is no memory, i.e., $\hat s_0^0=s_0^0$ and $\hat s_l^0=0$, $l\neq 0$. Noise, however, 
causes  information
to flow from $s_0^0$ to $\hat s_l^0$ for all $l$, so that the second
term in \eqref{eq:chain-rule} is non-zero. This term, which we call
the \emph{stochastic memory}, is ignored in our simulations. As a
consequence, the rates shown in Fig.~\ref{fig:rates} (a) are 
lower bounds to the corresponding maximum achievable rates.

From the conservation of energy, when the memory is large, $|\hat s_0^0|<|s_0^0|$, i.e., 
the received symbols cluster in a cloud that is closer to origin than
$s_0^0$. Limiting the detection at the RX to $\hat s_0^0$ results in an
output SNR loss. We observed that, as the input power 
is increased, the stochastic memory grows and the output SNR is decreased.
This is one reason that the NFDM lower bound in Fig. ~\ref{fig:rates}
(a)  saturates at high powers: we do not account for all received symbols. The stochastic memory exists in WDM with back-propagation as well.
Compared with the defocusing regime, continuous spectrum modulation in
the focusing regime produces signals
with larger peak-to-average ratio and stochastic memory.

\begin{figure}[t]
\centerline{
\begin{tabular}{ccc}
\includegraphics[width=0.27\textwidth]{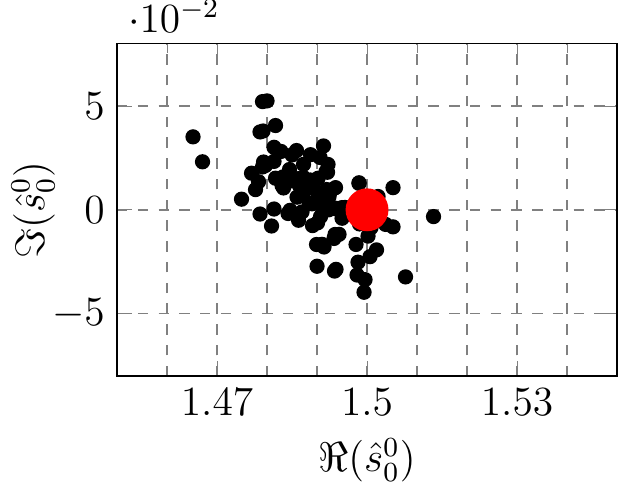}
&
\includegraphics[width=0.275\textwidth]{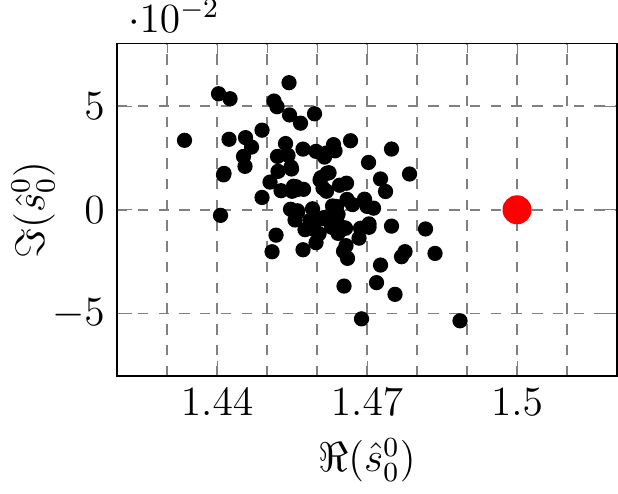}
&
\includegraphics[width=0.275\textwidth]{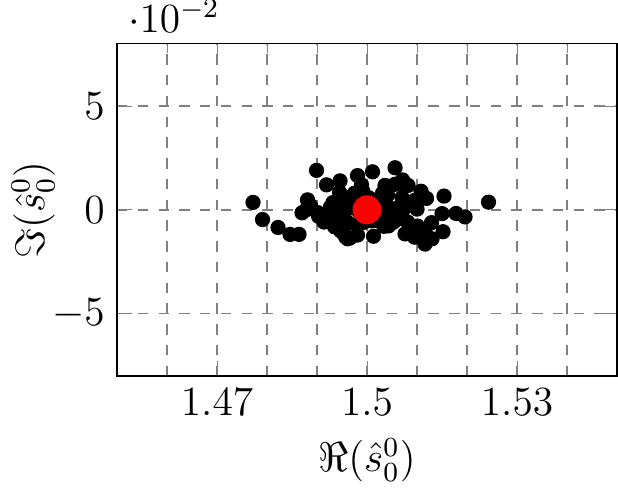}
\\
~~~~~~~~~~~~(a) &~~~~~~~~~~~~(b) & ~~~~~~~~~~~~~(c)
\end{tabular}
}
\caption{Transmit and received symbols in the (a) defocusing regime, (b) focusing regime,  and (c) the
AWGN channel. Here, $B=60$ GHz, $N_u=5$ and $s_0^0=0.5$. Transmit energy is the same in all cases.}
\end{figure}

\bibliographystyle{osameet2}
% \bibliography{Refs}

\begin{thebibliography}{1}
{\fontsize{8pt}{9.6pt}\selectfont

\bibitem{yousefi2016nfdm}
M.~I. Yousefi, et al., ``Linear and nonlinear frequency-division
  multiplexing,'' ar{X}iv:1603.04389 (2016).
 
\bibitem{part123}
M.~I. Yousefi, et al., ``Information transmission using the
  nonlinear {F}ourier transform, {P}art {I, II, III},'' IEEE Trans. Inf.
Theory \textbf{60}, 4312--4369 (2014). 

\bibitem{prilepsky2013}
J.~E. Prilepsky, et al., ``Nonlinear spectral management: Linearization
of the lossless fiber channel,'' Opt. Exp. \textbf{21}, 24344--24367 (2013).


\bibitem{le2014nis}
S. T. Le, et al., ``Nonlinear inverse synthesis for high spectral
efficiency transmission in optical fibers,''
Opt. Exp. \textbf{22}, 26720--26741 (2014).


\bibitem{wahls2015digital}
S.~Wahls, et al., ``Digital
  backpropagation in the nonlinear {F}ourier domain,'' in \ul{IEEE Int. Workshop Signal Process. Advances  
Wireless Commun.}, Stockholm, Sweden (2015), pp. 445--449.


\bibitem{aref2016ecoc}
V.~Aref, et al., ``Demonstration of fully nonlinear spectrum
  modulated system in the highly nonlinear optical transmission
  regime,'' in \ul{European Conf. Opt. Commun.}, D\"usseldorf, Germany (2016), pp. 19--21.

\bibitem{tavakkolnia2015sig}
I.~Tavakkolnia, et al., ``Signalling over nonlinear fibre-optic channels
  by utilizing both solitonic and radiative spectra,'' in \ul{European Conf. Networks and Commun.}, 
Paris, France (2015), pp. 103--107.

\bibitem{turitsyn2016nature}
S.~A. Derevyanko,  et al., ``Capacity estimates for
  optical transmission based on the nonlinear {F}ourier transform,''
  Nature Commun. \textbf{7}, 12710 (1--9), (2016).


}
\end{thebibliography}

\end{document}